\documentclass[conference]{IEEEtran}
\usepackage{myStyleIEEE}
\usepackage{ushort}
\IEEEoverridecommandlockouts

\def\R{\mathbb{R}}

\usepackage{siunitx}
\AtBeginDocument{\DeclareSIUnit{\MWh}{MWh}}
\AtBeginDocument{\DeclareSIUnit{\kWh}{kWh}}
\AtBeginDocument{\DeclareSIUnit{\voltampere}{VA}}

\usepackage{amsmath}
\usepackage{mathtools}
\usepackage{isomath}

\usepackage{eurosym}
\DeclareRobustCommand{\officialeuro}{%
  \ifmmode\expandafter\text\fi
  {\fontencoding{U}\fontfamily{eurosym}\selectfont e}}
  
\begin{document}

\title{Safe Reinforcement Learning for Strategic Bidding of Virtual Power Plants in Day-Ahead Markets}

\renewcommand{\theenumi}{\alph{enumi}}

\newcommand{\uros}[1]{\textcolor{magenta}{$\xrightarrow[]{\text{U}}$ #1}}
\newcommand{\justin}[1]{\textcolor{blue}{$\xrightarrow[]{\text{J}}$ #1}}
\newcommand{\ognjen}[1]{\textcolor{red}{$\xrightarrow[]{\text{O}}$ #1}}

\author{
\IEEEauthorblockN{Ognjen Stanojev\IEEEauthorrefmark{1}, Lesia Mitridati\IEEEauthorrefmark{2}, Riccardo de Nardis di Prata\IEEEauthorrefmark{1}, Gabriela Hug\IEEEauthorrefmark{1}}%
\IEEEauthorblockA{\IEEEauthorrefmark{1} EEH - Power Systems Laboratory, ETH Zurich, Switzerland } %
\IEEEauthorblockA{\IEEEauthorrefmark{2} Center for Electric Power and Energy, Technical University of Denmark, Kgs. Lyngby, Denmark} %
Emails: \{stanojev, hug\}@eeh.ee.ethz.ch, rprata@ethz.ch, lemitri@dtu.dk
\thanks{Research supported by NCCR Automation, a National Centre of Competence in Research, funded by the Swiss National Science Foundation (grant number 51NF40\_180545).}
}

\maketitle
\IEEEpeerreviewmaketitle

\begin{abstract}
This paper presents a novel safe reinforcement learning algorithm for strategic bidding of Virtual Power Plants (VPPs) in day-ahead electricity markets. The proposed algorithm utilizes the Deep Deterministic Policy Gradient (DDPG) method to learn competitive bidding policies without requiring an accurate market model. Furthermore, to account for the complex internal physical constraints of VPPs, we introduce two enhancements to the DDPG method. Firstly, a projection-based safety shield that restricts the agent's actions to the feasible space defined by the non-linear power flow equations and operating constraints of distributed energy resources is derived. Secondly, a penalty for the shield activation in the reward function that incentivizes the agent to learn a safer policy is introduced. A case study based on the IEEE 13-bus network demonstrates the effectiveness of the proposed approach in enabling the agent to learn a highly competitive, safe strategic policy.
\end{abstract}


\section{Introduction} \label{sec:intro}

Growing environmental concerns and advancements in communication and monitoring technologies have led to the increased deployment of Distributed Energy Resources (DERs) in power networks \cite{Parag2016ElectricityEra}, comprising renewable energy sources and \textit{prosumers}.
The market integration of these units is facilitated by their large-scale aggregation under financial entities, commonly known as Virtual Power Plants (VPPs), which have the capacity for trading in wholesale electricity markets \cite{VPPDef,Stanojev2022MultipleNetworks}. As a self-interested market participant, a VPP aims at maximizing its own profit generated by its market participation and the fulfillment of contractual obligations towards its internal customers. Concurrently, VPPs must ensure the safe operation of the underlying distribution network.

In electricity markets with imperfect competition, self-interested market participants, such as large-scale VPPs, may exercise market power by bidding strategically to influence the clearing prices and increase their profits \cite{Kirschen2018FundamentalsEd.}.
This bidding strategy has traditionally been modeled as a hierarchical optimization problem \cite{Kardakos2016OptimalApproach, yi2020bi}, in which the strategic market participant can anticipate the impact of its bids on the market clearing problem, provided that it has accurate information on all other parameters of the market clearing process.  
However, this approach has many limitations in practice, including the assumption of perfect information on other market participants' bidding strategies, leading to optimistic and unrealistic solutions of the market clearing. Furthermore, computational complexity is often an issue, even under limiting assumptions on the convexity of the market clearing problem \cite{Zemkohoo2020BilevelChallenges}.

The rapid progression in Reinforcement Learning (RL) has opened up new possibilities for market participants to devise sophisticated bidding strategies and learn to maximize their profits by interacting with the environment (i.e., market) \cite{Cao2020ReinforcementReview}.
One key advantage of model-free RL methods is that a strategic market participant (RL agent) does not require any prior knowledge of the market-clearing process or the other participants' strategies (environment).
However, traditional $Q$-learning-based RL algorithms struggle with large state-action spaces, as they rely on complex tables to approximate the value function for each state-action pair, leading to trade-offs between optimality, computation time, and memory allocation \cite{tellidou2006multi}. For this reason, their applicability in practice is limited. 

The above-mentioned scalability issues can be addressed by employing deep RL methods like the Deep Deterministic Policy Gradient (DDPG) algorithm \cite{Lillicrap2015ContinuousLearning}, which utilizes neural networks to extend the $Q$-learning capabilities to continuous state and action spaces.
The authors in \cite{Lin2020DeepEnergy,Ye2020DeepMarkets,zhu2022optimal} propose deep RL methods for the economic dispatch and market participation of DERs aggregated in a VPP. The main limitation of these works is that they fail to account for the complex internal physical constraints of large-scale VPPs, such as power generation limits and power flow constraints, in order to ensure a safe operation. In safety-critical infrastructures, such as distribution networks, this constitutes a significant limitation for the deployment of RL-based methods.
Consequently, it is necessary to investigate safe RL methods that can reliably ensure the fulfillment of these physical constraints.

Recent approaches in the field of safe RL can broadly be classified into two categories, providing improvements to (i) the reward function or (ii) the exploration phase of RL algorithms \cite{Garcia2015ALearning}. In the first category, a popular approach is to augment the reward function to account for constraint violations using the Lagrangian relaxation method \cite{yu2022towards}. However, such approaches may require ad hoc tuning of the constraint violation reward and may result in unsafe decisions during the exploration phase. In the second category, the safety of the decisions is promoted by offline (batch) learning to initialize the exploration \cite{lesage2022batch} or by the transfer of expert knowledge learned offline to guide the exploration \cite{grbic2020safer,thananjeyan2021recovery,geramifard2012practical}. Despite significant improvements, these approaches cannot provide safety guarantees and are not suitable for fully online learning.

Given the aforementioned challenges, this paper focuses on the strategic bidding problem of VPPs in day-ahead electricity markets and introduces a new safe RL algorithm in this context. The proposed approach employs the DDPG algorithm to facilitate the learning of complex policies that operate in continuous action and state spaces. To enhance safety, we combine (i) a \textit{safety shield} that restricts the agent's actions to a safe space defined by the non-linear power flow equations and DER operating constraints and (ii) a penalty for shield activation in the reward function that incentivizes the agent to learn a safer policy. The projection-based safety shield acts as a protective measure that prohibits the agent from making unsafe decisions and instead selects the closest safe action, extending the works in \cite{KPWabersich,SafeProjection,Badoual2021AStorage}. The proposed approach provides flexibility in the design of the agent and the representation of its economic and physical aspects. Numerical results based on the IEEE 13-bus network demonstrate the ability of the developed agent to devise a highly competitive and safe bidding strategy.

The subsequent sections of this paper are organized as follows. Section~\ref{sec:vpp_market_models} provides an overview of VPP modeling and introduces the market clearing problem under consideration. In Section~\ref{sec:OPF}, we introduce the proposed bidding scheme, delineating the RL problem formulation and the derivation of the safety-enforcing optimization problem. Section~\ref{sec:res} presents the case studies conducted to validate and assess the bidding scheme's performance, while Section~\ref{sec:concl} concludes the paper.

\section{Virtual Power Plant and Market Models} \label{sec:vpp_market_models}

The VPP under study is assumed to serve load and operate a multitude of heterogeneous DERs within a distribution network, namely, renewable energy sources (wind or PV), energy storage units, and conventional generation such as diesel generators. The underlying distribution network is considered to be radial and balanced, and can thus be represented by a tree graph $\mathcal{G}(\mathcal{N},\mathcal{E})$, with $\mathcal{N} \coloneqq \{0,1,\dots,n\}$ denoting the set of network nodes including the substation node $0$, and $\mathcal{E} \subseteq \mathcal{N}\times\mathcal{N}$ representing the set of $n$ network branches. A single Point of Common Coupling (PCC) of the VPP to the rest of the system at node $0$ is assumed. Loads are considered to be located at nodes $\mathcal{L}\subseteq\mathcal{N}$, Renewable Energy Sources (RESs) at nodes $\mathcal{R}\subseteq \mathcal{N}$, Battery Energy Storage Systems (BESSs) at nodes $\mathcal{B}\subseteq \mathcal{N}$, and conventional generators at nodes $\mathcal{D}\subseteq \mathcal{N}$. Without loss of generality, each node is assumed to contain a single DER type, i.e., $\mathcal{R}\cap\mathcal{B}\cap\mathcal{D}=\varnothing$.

\subsection{Grid Constraints - Flow and Voltage Deviation Limits} \label{sec:hnet}
For every bus $i\in\mathcal{N}$, let $v_i\in\R_{\geq 0}$ denote the voltage magnitude, and $p_{i}\in\R$ and $q_{i}\in\R$ represent the active and reactive nodal power injections. For each branch $(i,j)\in\mathcal{E}$, let $r_{ij}\in\R_{\geq 0}$ and $x_{ij}\in\R$ denote its respective resistance and reactance values, $P_{ij}\in\R$ and $Q_{ij}\in\R$ denote the real and reactive power flow along the branch, and $i_{ij}\in\R_{\geq 0}$ represent the corresponding branch current magnitude. The distribution grid is modelled using the DistFlow equations \cite{Baran1989}, written recursively for every line $(i,j)\in\mathcal{E}$ as:
\begin{subequations} \label{eq:distFlow}
\begin{align}
    P_{ij}  &= \sum_{k\in\mathcal{N}_j} P_{jk} - p_{j} + r_{ij}i_{ij}^2, \label{eq:p_distFlow} \\
    Q_{ij}  &= \sum_{k\in\mathcal{N}_j} Q_{jk} - q_{j} + x_{ij}i_{ij}^2, \label{eq:q_distFlow} \\
    v_i^2 - v_j^2 &= 2(r_{ij}P_{ij}+x_{ij}Q_{ij}) - (r_{ij}^2+x_{ij}^2)i_{ij}^2, \label{eq:v_distFlow}
\end{align}
\end{subequations}
with the branch currents computed as $i_{ij}^2=(P_{ij}^2+Q_{ij}^2)/v_i^2$, and $\mathcal{N}_j$ denoting the set of child nodes of bus $j$.
 
To respect power quality standards and preserve the health of assets within the VPP, network constraints in the form of voltage deviations and line power limits need to be satisfied:
\begin{alignat}{2}
     S_{ij} &\leq S^{\mathrm{max}}_{ij},\qquad\quad &&\forall(i,j)\in\mathcal{E}, \label{eq:s_limit} \\
     v^{\mathrm{min}}_i \leq v_i &\leq v^{\mathrm{max}}_i, \qquad\quad &&\forall i\in\mathcal{N}, \label{eq:v_limit}
\end{alignat}
where $S_{ij}^2=P_{ij}^2+Q_{ij}^2$ is the apparent power flowing through line $(i,j)$, and $\mathrm{min}$ and $\mathrm{max}$ superscripts indicate respectively the minimum and the maximum allowable values. 

\subsection{DERs Constraints - Individual Capability Curves} \label{sec:gen_constr}
The active and reactive power outputs of the DERs within the VPP are additionally limited by their internal constraints. These constraints are typically defined for each unit $i$ as sets of allowable operating points $(p_{i},q_{i})\in\mathcal{X}_{i}\subseteq\R^2$, called capability curves. The capability curve of a renewable generator $i\in\mathcal{R}$ can be defined by the following set: 
\begin{equation*}
\begin{split}
    \mathcal{X}_i \coloneqq \{(p_{i},q_{i})\in\R^2: 0\leq p_{i} \leq p^\mathrm{res}_{i}, s_i\leq s^\mathrm{max}_i,  \cos{\phi}\geq a\},
\end{split}
\end{equation*}
where $s^\mathrm{max}_{i}\in\R_{\geq0}$ denotes the apparent power limit and $p^\mathrm{res}_{i}\in\R_{\geq0}$ represents the irradiation/wind speed dependent active power upper limit, while
the lower limit is zero due to the possibility of curtailment. Due to its dependency on exogenous variables, the upper limit $p^\mathrm{res}_i$ is modeled as a random variable with an appropriate probability distribution function derived from the historical data. Furthermore, power quality standards commonly require the power factor $\cos{\phi}$ to be greater than a prescribed value $a\in[0,1]$.

The BESS capability curve is assumed to allow four-quadrant operation and is defined for each unit $i\in\mathcal{B}$ by:
\begin{align*}
    \mathcal{X}_i \coloneqq \{(p_{i},q_{i})\in\R^2: -&p^\mathrm{max}_{i} \leq p_{i} \leq p^\mathrm{max}_{i}, s_i\leq s^\mathrm{max}_i, \\
    &\chi^\mathrm{min}_{i} \leq \chi_{i} \leq \chi^\mathrm{max}_{i}
    \},
\end{align*}
with $p^\mathrm{max}_{i}\in\R_{\geq0}$ denoting the maximum charging/discharging power, and $s^\mathrm{max}_{i}\in\R_{\geq0}$ representing the apparent power limit. Storage energy capacity limits constrain the state-of-energy $\chi\in\R_{\geq0}$ between  $\chi^\mathrm{min}\in\R_{\geq0}$ and $\chi^\mathrm{max}\in\R_{\geq0}$. 

Operation of each conventional generator $i\in\mathcal{D}$ is constrained by stator current and low load limits, which impose the minimum $p_i^\mathrm{min}\in\R_{\geq0}$ and the maximum $p_i^\mathrm{max}\in\R_{\geq0}$ active power output, as well as the maximum apparent power $s_i^\mathrm{max}\in\R_{\geq0}$ limits, that is,
\begin{equation*}
    \mathcal{X}_i \coloneqq \{(p_{i},q_{i})\in\R^2: p^\mathrm{min}_i \leq p_{i} \leq p_i^\mathrm{max},\,\, s_i\leq s^\mathrm{max}_i\}.
\end{equation*}

Loads are assumed to be non-flexible. Therefore, we have $\mathcal{X}_i \coloneqq \{(p_{i},q_{i})\in\R^2: p_i=-p_i^\mathrm{load},q_i=-q_i^\mathrm{load}\}$ for load $i\in\mathcal{L}$, where $p_i^\mathrm{load}\in\R_{\geq0}$ and $q_i^\mathrm{load}\in\R_{\geq0}$ are the constant active and reactive power consumptions. Finally, the buses that are not hosting any generation or consumption element have zero injections, i.e., $p_i=q_i=0,\forall i\in\mathcal{N}\setminus(\mathcal{L} \cup \mathcal{R} \cup \mathcal{D} \cup \mathcal{B})$.

\subsection{Day-Ahead Market Clearing Model}
The day-ahead electricity market is operated through a blind auction which takes place once a day, when each market participant submits independent price-quantity bids for each hour of the following day. We distinguish between \textit{supply} bids defined by $(\gamma^s_n,p^{s,\mathrm{bid}}_n)$ for each supplier $n\in\mathcal{S}$, where $\gamma^s_n$ denotes the price offer for the quantity $p^{s,\mathrm{bid}}_n$; and \textit{demand} bids defined by $(\gamma^d_j,p^{d,\mathrm{bid}}_j)$ for each customer $j\in\mathcal{C}$, where $\gamma^d_j$ denotes the price customer $j$ is willing to pay for the quantity $p^{d,\mathrm{bid}}_j$. The market clearing process is modeled by the following optimization problem:
\begin{subequations}\label{eq:market_clearing}
\begin{alignat}{3}
 ({\mathcal{P}_1})\quad& \underset{p^s_n, p^d_j}{\max}\quad &&  \sum_{j\in\mathcal{C}} \gamma^d_jp^d_j - \sum_{n\in\mathcal{S}} \gamma^s_np^s_n \label{eq:SW} \\
& \,\,\textrm{s.t.} && \sum_{j\in\mathcal{C}} p^d_j = \sum_{n\in\mathcal{S}} p^s_n, \label{eq:mc_balance}\\
& && 0\leq p^d_j \leq p^{d,\mathrm{bid}}_j, \forall j\in\mathcal{C}, \label{eq:mc_demand} \\
& && 0 \leq p^s_n \leq p^{s,\mathrm{bid}}_n, \forall n\in\mathcal{S},\label{eq:mc_supply}
\end{alignat}
\end{subequations}
aiming at maximizing the social welfare \eqref{eq:SW}, subject to constraints on the power balance in the transmission grid \eqref{eq:mc_balance} and limits on the cleared quantities $(p^s_n,p^d_j)$ \eqref{eq:mc_demand}-\eqref{eq:mc_supply}. All the market participants are paid the uniform Market Clearing Price (MCP), which is computed as the dual variable of \eqref{eq:mc_balance}.

\section{Safe RL-based Bidding Strategy} \label{sec:OPF}
\subsection{Agent-Environment Interaction}
\begin{figure}[!t] 
\centering
    \includegraphics[scale=0.6125]{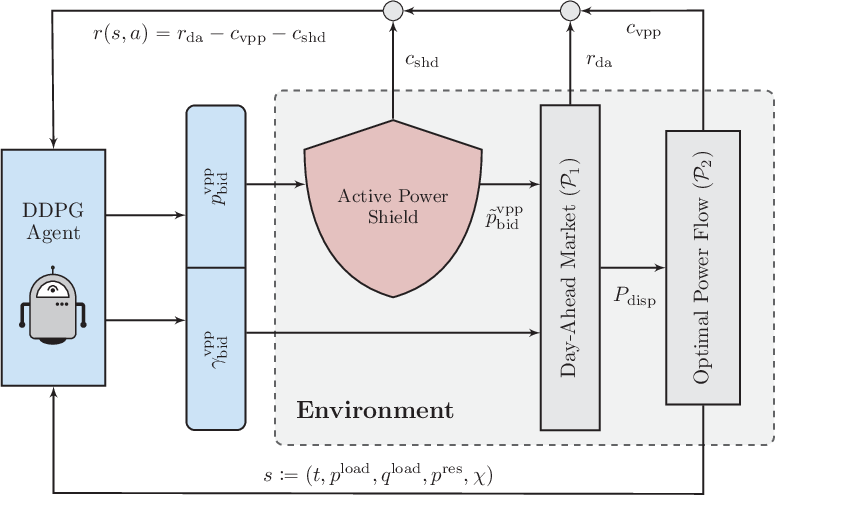}
 \caption{The proposed agent-environment interaction scheme.}
\label{fig:reduced_model}
\end{figure}

The proposed bidding process of the strategic VPP in the day-ahead market is presented in Figure~\ref{fig:reduced_model}. The bidding decisions are determined by an RL-based agent, designed as a single-agent actor-critic model exploiting the DDPG algorithm. The agent is in control of generating two values, namely the bidding quantity $p_\mathrm{bid}^\mathrm{vpp}\in\R$ and the bidding price $\gamma^\mathrm{bid}_\mathrm{vpp}\in\R_{\geq0}$, collected in the action vector $a \coloneqq (\gamma^\mathrm{bid}_\mathrm{vpp},p_\mathrm{bid}^\mathrm{vpp})$.

The agent interacts with the environment composed of the \textit{shield}, the market clearing engine, and an Optimal Power Flow (OPF) module. The generated bid is first passed to the \textit{shield} that performs a nonlinear projection $\Tilde{p}_\mathrm{bid}^\mathrm{vpp} = \proj_\mathcal{A}(p_\mathrm{bid}^\mathrm{vpp})$ on a safe set $\mathcal{A}$, which captures the feasible operating space of the VPP. The corrected bidding quantity $\Tilde{p}_\mathrm{bid}^\mathrm{vpp}$ is submitted to the market which is then cleared according to the procedure described in \eqref{eq:market_clearing}. After the market clearing, an OPF algorithm is run to dispatch the units within the VPP to respect the cleared capacity. The environment state is defined by $s \coloneqq (t,p^\mathrm{load},q^\mathrm{load},p^\mathrm{res},\chi)$. Using the hour of the day (i.e., the time step $t$) as a state variable the agent attempts to learn daily price patterns in order to place better bids. To understand how much power it needs to buy, or how much power generation it has left to sell, it is necessary to pass the expected load values $(p^\mathrm{load},q^\mathrm{load})$ and renewable generation $p^\mathrm{res}$ output as state variables. Lastly, the state-of-energy vector $\chi$ indicates the VPP's storage capacity available to the agent. Note that time indexing of variables is omitted. Unless otherwise specified, each variable refers to the current time step $t$ hereafter.

The reward $r(s,a) = r_\mathrm{da} - c_\mathrm{vpp} -c_\mathrm{shd}$ consists of three different components, the market profit $r_\mathrm{da}$, the VPP internal cost $c_\mathrm{vpp}$, and the shield activation cost $c_\mathrm{shd}$. The first component is the profit that the RL agent generates by bidding in the market and is computed as $r_\mathrm{da}=P_\mathrm{disp}\cdot\mathrm{MCP}$. Since the day-ahead market is a uniform price auction, the dispatched (cleared) quantity is multiplied by the MCP. The second component $c_\mathrm{vpp}$ reflects the internal financial activities of the VPP. The internal costs are calculated by subtracting the load supply revenues from the power generation costs:
\begin{equation}
    c_{\mathrm{vpp}} =  \sum_{i \in {\mathcal{R}\cup\mathcal{D}}} p_i \gamma_i - \sum_{n \in\mathcal{L}} p_n \gamma_n.
\end{equation}
Lastly, the shield-activation penalty, which quantifies how much the shield had to intervene on the original bid in order to make it feasible, is defined as
$c_{\mathrm{shd}} = \epsilon \| \Tilde{p}_\mathrm{bid}^\mathrm{vpp} - p_\mathrm{bid}^\mathrm{vpp} \|$.
The non-negative weight $\epsilon\geq0$ is a tuning parameter.

\subsection{Deep Deterministic Policy Gradient Algorithm} \label{subsec:ddpg}
Deep Deterministic Policy Gradient is an actor-critic, model-free RL algorithm \cite{Lillicrap2015ContinuousLearning} suitable for continuous state and action space problems such as the considered VPP bidding problem. The actor-critic methods employ a policy function and an action-value function. The policy function acts as an actor, generating an action $a$ given a state observation $s$, whereas the action-value function acts as a critic, which takes the state-action pair $(s,a)$ as the input and outputs the $Q$-value associated with this pair. In the DDPG algorithm, two neural networks are used to approximate the action-value function and the policy function: the critic network $Q(s,a\,\lvert\, \theta^{Q})$ and the actor network $\mu(s\,\lvert\, \theta^{\mu})$, respectively, with $\theta^{Q}$ and $\theta^{\mu}$ being the parameters of the respective networks \cite{Stanojev2020}. 

Furthermore, two additional networks, the target actor network $\mu'(s\,\lvert\,\theta^{\mu'})$ and the target critic network $Q'(s,a\,\lvert\,\theta^{Q'})$ are introduced as time-delayed copies of the actor and the critic: 
\begin{equation*} \label{eq:target_net_updates}
    \theta^{Q'} \mapsfrom \tau \theta^{Q} + (1-\tau) \theta^{Q'}, \quad
    \theta^{\mu'} \mapsfrom \tau \theta^{\mu} + (1-\tau) \theta^{\mu'},
\end{equation*}
with the tracking rate defined by the soft update coefficient $\tau \ll 1$. These additional networks are used for rendering the learning process more stable by creating the labels for training of the original networks, as will be shown in the following.

The training process is conducted by running a predefined number of episodes, each comprising a series of steps in which the agent interacts with the environment. During the training, the agent explores the environment by allowing less favorable actions with respect to the current knowledge. The exploration is achieved using Gaussian noise $\mathcal{N}(0,\sigma(m))$ with zero mean and an exponentially decreasing variance $\sigma(m)$ defined by 
\begin{equation*}\label{eq:gauss_noise}
    \sigma(m) = z_\mathrm{i} \exp\Big({-m \frac{\ln(\frac{z_\mathrm{i}}{z_\mathrm{f}})}{m_\mathrm{tot}}}\Big),
\end{equation*}
where $m$ is the episode number, $m_\mathrm{tot}$ is the total number of episodes used for training, $z_\mathrm{i}$ is the desired initial noise, $z_\mathrm{f}$ is the desired final noise. The noise sampled from the Gaussian process is superposed on the actor's output $a_m = \mu(s_m \,\lvert\, \theta^{\mu} )+\mathcal{N}(0,\sigma(m))$. After each interaction, the tuple $(s_m,a_m,r_m,s_{m+1})$ is stored into the experience replay memory, from which the minibatches (i.i.d. sets of samples) are selected for training of the original networks. The replay memory is a technique used to speed up and enhance learning.

Finally, we present the update rules for the actor and critic networks. The critic network is updated by minimizing the loss function, averaged over $N$ samples in the minibatch:
\begin{equation*} \label{eq:critic_loss}
L(\theta^{Q}) = \frac{1}{N} \sum_{i=1}^{N}(y_{i} - Q(s_i,a_i\,\lvert\, \theta^{Q}))^2.
\end{equation*}
The label $y_{i}$ for the $i$-th sample in the minibatch is calculated as a sum of the immediate reward received in that sample and the expected $Q$-function value of the next state $s_i'$, determined by the target actor and critic networks, i.e.,
\begin{equation*} \label{eq:critic_label}
y_{i} = r_{i} + \varepsilon Q'\left(s_{i}', \mu'(s_{i}' \,\lvert\, \theta^{\mu'}) \,\lvert\, \theta^{Q'}\right),
\end{equation*}
where $\varepsilon\geq0$ is the discount factor.
The actor network can be improved by maximizing the policy score function defined by
\begin{equation*} \label{eq:policy_score}
    J(\theta^\mu)=\mathop{\mathbb{E}} \big[Q(s,a \,\lvert\, \theta^{Q}) \,\lvert\,  _{s=s_{i},a=\mu(s_{i})}],
\end{equation*}
which evaluates performance of policy $\mu(\cdot\,\lvert\,\theta^\mu)$ for each sample in the minibatch. To maximize the score function, gradient ascent is thus applied to the actor network. This involves approximating the gradient by the average value of the policy score function gradients across the minibatch:
\begin{equation*}
\nabla_{\theta^{\mu}} J \approx \frac{1}{N} \sum_{i=1}^{N} (\nabla_{a} Q(s_i,a \,\lvert\, \theta^{Q})\,\lvert\,  _{a=\mu(s_{i})} \nabla_{\theta^{\mu}} \mu(s_i \,\lvert\, \theta^{\mu} )).
\end{equation*}

\subsection{Safe RL via Projection on the Feasible Space}
While the described DDPG algorithm has previously been shown to converge to effective bidding strategies in an electricity market framework \cite{Ye2020DeepMarkets}, there is yet no guarantee that the bidding decisions are feasible with respect to the VPP constraints. Namely, the submitted quantity might exceed the aggregate capacity of the VPP units or cause violations in grid voltages and currents. 
To alleviate this issue, the actions generated by the learned policy are projected into a safe set. The projection thus operates as a \textit{shield} that prevents RL from taking unsafe decisions and adopts the closest safe action to the generated unsafe RL decision, in the $\ell_2$-norm sense.

As safety limitations, in this paper we consider the VPP constraints introduced in Sec.~\ref{sec:hnet} and Sec.~\ref{sec:gen_constr} that must be respected for each submitted bid. This set of constraints is denoted by $\mathcal{A}$. To perform the projection of a generated bid onto $\mathcal{A}$, the following optimization problem is solved:
\begin{equation}\label{eq:shield_proj}
\begin{aligned} 
    \Tilde{p}_\mathrm{bid}^\mathrm{vpp} = \argmin_{u} \, & \frac{1}{2}\| u - p_\mathrm{bid}^\mathrm{vpp} \|^2_2 \\
    \textrm{s.t.} \quad & \eqref{eq:distFlow}, \eqref{eq:s_limit},\eqref{eq:v_limit},\\
    & (p_i,q_i)\in\mathcal{X}_i,\quad\forall i\in\mathcal{N},\\
    & P_{01} = - u,
\end{aligned}
\end{equation}
where the final equality constraint is a boundary condition expressing that $\Tilde{p}_\mathrm{bid}^\mathrm{vpp}$ is to be dispatched at the PCC.

It should be noted that this procedure does not provide a guarantee that the projected action is the most optimal within the set of feasible actions $\mathcal{A}$. However, through the use of a reward system that combines the profit generated by the feasible action $\tilde{p}_\mathrm{bid}^\mathrm{vpp}$ with a penalty for actions outside of $\mathcal{A}$, the RL agent can learn to restrict its actions to within the safe set $\mathcal{A}$. Thus, as the learning process continues, the agent will eventually converge to locally or globally optimal safe actions.

\subsection{Optimal Power Flow Dispatch}
The OPF procedure is used to determine the operating points of all the generators inside the VPP. Additionally, the results of the OPF can indicate the Production Marginal Cost (PMC) of the VPP, i.e., the production cost of the last internal generator dispatched. The considered OPF problem is formulated as:
\begin{equation} \label{eq:opf_nnl}
\begin{aligned}
    ({\mathcal{P}_2})\quad\min_{p_i,q_i} \quad & \sum_{i \in {\mathcal{R}\cup\mathcal{D}}} p_i \gamma_i-\sum_{n \in\mathcal{L} } p_n \gamma_n\\ 
    \textrm{s.t.} \quad & \eqref{eq:distFlow}, \eqref{eq:s_limit},\eqref{eq:v_limit},\\
    & (p_i,q_i)\in\mathcal{X}_i,\quad\forall i\in\mathcal{N},\\
    & P_{01} = - P_\mathrm{disp}.
\end{aligned}
\end{equation}
The value of the objective function is the previously defined internal VPP cost $c_\mathrm{vpp}$.
On the other hand, the production marginal cost of the VPP is equal to the cost difference of producing one more kWh of power. Note that the above problem \eqref{eq:opf_nnl} is a nonlinear, nonconvex optimization problem due to the inclusion of DistFlow constraints \eqref{eq:distFlow}. To enhance the computational tractability of this problem, one may need to resort to relaxations or approximations. Nonetheless, the pursuit of such techniques lies beyond the scope of this paper.

\section{Numerical Results}\label{sec:res}

\subsection{Case Study Setup}
\begin{figure}[!b]
    \centering
    \includegraphics[scale=0.65]{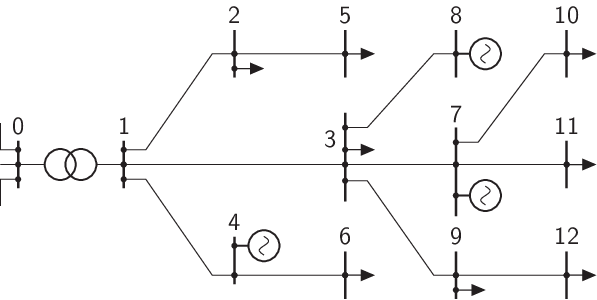}
    \caption{Single line diagram of the modified IEEE 13-bus system, with DERs placed according to the basecase configuration.}
    \label{fig:IEEE13busmodel}
\end{figure}

The algorithm developed in this paper is tested on a modified, single-phase version of the IEEE 13-bus distribution feeder \cite{TestFeeders2018}, shown in Fig.~\ref{fig:IEEE13busmodel}. A number of DERs have been introduced to transform the original distribution grid into a VPP. 
In order to construct a system with realistic load data, the dataset provided in \cite{Murray2017AnStudy} is used, where load from 20 households has been measured and stored continuously over a period of two years. The household load power values are normalized and adjusted to have the nominal capacity equal to the specifications of the IEEE 13-bus distribution feeder. All the loads $i\in\mathcal{L}$ are paying a fixed price of $\gamma_i=\SI[per-mode=symbol]{0.5}{\EUR\per\kWh}$ to the VPP owner for the provided supply.

While the network physical constraints (i.e., bus voltage and line thermal limits) are provided by the IEEE 13-bus specifications, the DERs placement and capacity are customized, and 3 configurations are considered, as described below:
\begin{itemize}
    \item \textbf{Basecase}: Only conventional dispatchable generators are considered, such that $\mathcal{D}\coloneqq\{4,7,8\},\mathcal{R}=\varnothing,\mathcal{B}=\varnothing$. The generators adopt the following parameterization: $s^\mathrm{max}_4=s^\mathrm{max}_7=\SI{500}{\kilo\voltampere},s^\mathrm{max}_8=\SI{950}{\kilo\voltampere}$, the minimum power limits are set to zero $p^\mathrm{min}_i=0,\forall i\in\mathcal{D}$, and the maximum power limits match the apparent power limits $p^\mathrm{max}_i=s^\mathrm{max}_i,\forall i\in\mathcal{D}$. The generation cost coefficients are $\gamma_4=\SI[per-mode=symbol]{4}{\EUR\per\kWh}, \gamma_7=\SI[per-mode=symbol]{4.5}{\EUR\per\kWh}$ and $\gamma_8=\SI[per-mode=symbol]{5}{\EUR\per\kWh}$. When not stated otherwise, this model is used in simulations. 
    \item \textbf{Renew1}: The conventional generator at bus 8 in the Basecase configuration is substituted by a RES, with the same nominal capacity and zero production cost, such that $\mathcal{D}=\{4,7\},\mathcal{R}=\{8\},\mathcal{B}=\varnothing$.
    \item \textbf{Renew2}: All conventional generators in the Basecase configuration are replaced by RESs with the same nominal capacities and zero production costs, such that $\mathcal{D}=\varnothing,\mathcal{R}=\{4,7,8\},\mathcal{B}=\varnothing$.
    \item \textbf{Bat1}: A BESS is connected at bus 1, with the following parameterization: $\chi_1^\mathrm{min}=0, \chi_1^\mathrm{max}=\SI{4800}{\kWh}, p^\mathrm{max}_1=\SI{500}{\kilo\watt}, s^\mathrm{max}_1=\SI{500}{\kilo\voltampere}$. All other generators from the Basecase configuration remain unchanged, such that $\mathcal{D}\coloneqq\{4,7,8\},\mathcal{R}=\varnothing,\mathcal{B}=\{1\}$. The battery cost factor is 0.
\end{itemize}

The implementation of the DDPG-based RL agent is briefly discussed below. Both the actor and the critic network have two hidden layers with 512 neurons that employ the Rectified Linear Unit (ReLU) activation function. The output layer of the actor employs a sigmoid function to bound the actions. The constant for the soft update of the target networks is $\tau=0.005$. We train the RL agent with a minibatch size of $10^6$ and the following exploration noise parameters: $z_\mathrm{i}=0.5, z_\mathrm{f} = 0.1$ and $m_\mathrm{tot}=100$. The \text{{\fontfamily{qcr}\selectfont Adam}} optimizer is used to find the neural network weights with a learning rate $\alpha=0.001$ for the actor and $\beta=0.002$ for the critic. Three potential designs of the reward function and the shield are considered to examine the benefits and performance of the safety layer:
\begin{enumerate}
    \item Unsafe Reinforcement Learning (\textbf{uRL}): The VPP constraints are not taken into account, and the shield is bypassed. The reward consists only of day-ahead market revenues, more precisely $r(s,a) = r_\mathrm{da}$.
    \item Shielded Reinforcement Learning (\textbf{shRL}): The reward is designed as $r(s,a) = r_\mathrm{da}-c_\mathrm{vpp}$, with the constraints of the VPP taken into account via the active power shield \eqref{eq:shield_proj}. However, no negative reward is passed to the agent for the shield activation.
    \item Safe Reinforcement Learning (\textbf{sRL}): In this case, the shield is active and the agent receives a negative reward for its activation, i.e., $r(s,a) = r_\mathrm{da} - c_\mathrm{vpp} - c_\mathrm{shd}$. Here, the goal is to train an agent that can place strategic bids without violating the physical limits of the VPP.
\end{enumerate}

\begin{table}[!b]
\renewcommand{\arraystretch}{1}
\caption{Performance of the three different RL models, with all values given in \EUR{}/day.}
\label{tab:model_comparisons}
\noindent
\centering
    \begin{minipage}{\linewidth} 
    \begin{center}
\scalebox{0.9}{%
    {\setlength{\extrarowheight}{.15em}\tabcolsep=4pt
    \begin{tabular}{c||c c c | c | c}
    \toprule
    Model & Day-Ahead & Balancing Cost & Net Market Profit & $c_\mathrm{vpp}$ & $c_\mathrm{sh}$ \\
    \hline
    uRL & 162819.1 & 67163.8  & 95655.3 & - & -\\
    \arrayrulecolor{black!30}\cline{1-4}
    shRL & 157970.1 & 0 & 157970.1 & 93927.46 & -\\
    \arrayrulecolor{black!30}\cline{1-4}
    sRL & 154594.4 & 0 & 154594.4 & 93052.12 & 10E-05\\
    \arrayrulecolor{black}\bottomrule
    \end{tabular}}
    }
    \end{center}
    \end{minipage}
\end{table}
\subsection{Safe vs. Unsafe RL-based VPP Bidding Strategies}
\begin{figure}[!t]
    \centering
    \includegraphics[scale=1.125]{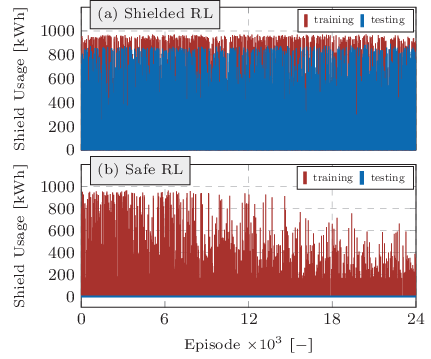}
    \caption{Shield activation during training and testing for (a) Shielded RL design (top) (b) Safe RL design (bottom).}
    \label{fig:shield_activation}
\end{figure}

The three different models listed above are evaluated on the Basecase scenario by assessing the average reward generated during the algorithm testing phase, which comprises $500$ episodes consisting of $24$ steps (hours). Table~\ref{tab:model_comparisons} showcases the performance of these models. It can be observed that the uRL model achieves the highest market revenues. However, as it by definition does not take into account the physical constraints of the grid, it often bids a higher quantity than the VPP can physically provide. Hence, it is obliged to buy balancing power, which has a higher cost than the market price. In this work, the balancing price is assumed to be $20\%$ higher than the MCP. Therefore, the total profit generated by the uRL decreases. The shRL and sRL models achieve similar market profits, with sRL sacrificing some performance for safety. The shield component of the reward in sRL is negligible, which signals that the RL algorithm has learned how to take actions within the physical boundaries imposed by the nature of the VPP. In Fig.~\ref{fig:shield_activation}, it can be seen that during the training phase of sRL the shield is activated very often, while in the testing phase it is not needed anymore. This is clearly not the case for shielded RL, as showcased in the same figure. The lower shield activation observed during the testing phase in the top plot can be attributed to a decreased level of exploration noise.

\subsection{Bidding Strategy Analysis}
\begin{figure}[!b]
    \centering
    \includegraphics[scale=0.675]{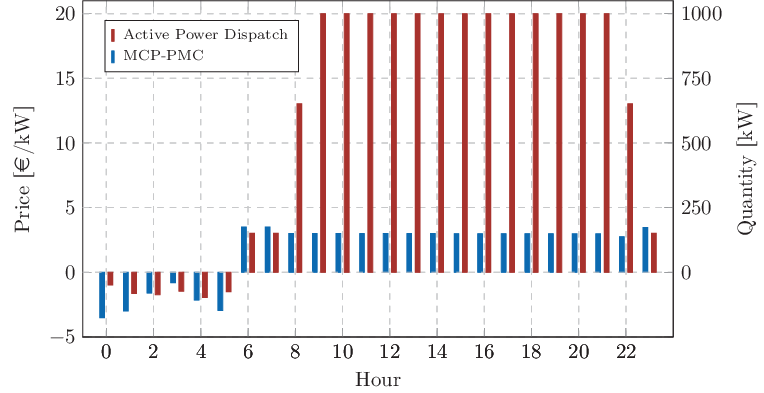}
    \caption{Bidding behavior of the safe RL agent over 24 hours.}
    \label{fig:behaviour_opt2}
\end{figure}

In this section, we analyze the bidding strategies that the sRL agent adopts in the basecase scenario to achieve higher profits in the day-ahead market, as showcased in Figure \ref{fig:behaviour_opt2}.
The VPP has the option to both buy and sell power on the day-ahead market, depending on the internally available generation and the consumption requirements of its customers (inflexible loads at nodes $i\in\mathcal{L}$). 
The VPP is expected to dispatch internal generators whose production cost is lower than the MCP, and to buy the potential remaining energy needed to supply its loads in the day-ahead market. This strategy can clearly be observed in several hours in Fig.~\ref{fig:behaviour_opt2}. In the first 6 hours of the day, the MCP is lower than the PMC of the VPP. Consequently, the VPP is buying power from the exchange. The amount of power varies hour by hour, as the VPP in most cases buys the exact amount needed to supply its loads. In case there is a generator with a unit PMC lower than the MCP, the amount of power bought is lower. During the other hours, the MCP is higher than the PMC, and hence, selling power becomes profitable. The VPP still provides the necessary power to its loads and generates extra power to sell to the market. The amount of power sold is limited either by its physical constraints
or by the demand on the market.

Besides learning to bid based on the difference between the MCP and the PMC of the VPP, the RL agent also learns to act as a price maker in the market. 
By learning the expected MCP and the bids supporting it, the agent can place bids between the expected MCP and the next rival participant's bid in order to increase the MCP and maximize its profits. This process is of a delicate nature, as a bid that is too high might become out-of-the-money, leading to the VPP not being cleared.

\begin{figure}[!t]
    \centering
        \includegraphics[scale=1.125]{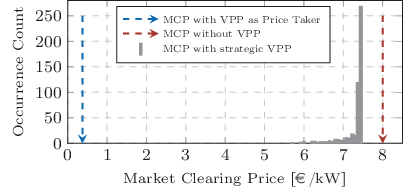}
    \caption{MCP with strategic VPP (RL agent), without VPP and with price-taker VPP.}
    \label{fig:behaviour_VPP_pricemaker}
\end{figure}

The bidding behavior of the RL agent at hour 22 and its impact on the MCP is illustrated in Fig.~\ref{fig:behaviour_VPP_pricemaker}. It suggests that the VPP acts as a price maker in this hour. Indeed, here, the range of feasible bids for the VPP is $\SI[per-mode=symbol]{0.5}{\EUR\per\kilo\watt}$ and $\SI[per-mode=symbol]{8}{\EUR\per\kilo\watt}$, as bidding above $\SI[per-mode=symbol]{8}{\EUR\per\kilo\watt}$ would get the VPP out of the merit order. A price maker VPP with perfect information on the rival participants' bids would bid as close as possible to the upper bound of this range, i.e., at $\SI[per-mode=symbol]{8}{\EUR\per\kilo\watt}$. We observe that by learning to increase its price bid within this competitive range, the RL agent manages to influence the MCP for most of the episodes, considerably increasing its net profit.

\subsection{Impact of Renewable Energy Sources and Batteries}
In this section, we analyze quantitatively and qualitatively the impact of the introduction of RESs and BESSs into the VPP in terms of the bidding behavior of the RL agent and the total average net profits of the VPP, as summarized in Table~\ref{tab:res_bess_comparison} for the different DERs configurations considered. The fields ``w FC'' and ``w/o FC'' indicate whether the renewable forecast is included in the agent state or not. Note that a different load curve compared to the previous subsections is used. 
\begin{table}[!b]
\renewcommand{\arraystretch}{1}
\caption{Performance of the considered DER configurations.}
\label{tab:res_bess_comparison}
\noindent
\centering
    \begin{minipage}{\linewidth} 
    \begin{center}
\scalebox{0.9}{%
    {\setlength{\extrarowheight}{.15em}\tabcolsep=4pt
    \begin{tabular}{c||c |c c c c| c}
    \toprule
    \multirow{2}{*}{Configuration} & \multirow{2}{*}{\textbf{Basecase}} & \multicolumn{2}{c}{\textbf{Renew1}} & \multicolumn{2}{c|}{\textbf{Renew2}} & \multirow{2}{*}{\textbf{Bat1}} \\
            \cline{3-6}
             & & w FC & w/o FC & w FC & w/o FC & \\
    \hline
    \begin{tabular}{c} Profit $[\EUR{}/\mathrm{day}]$ \end{tabular} & 44558 & 72983 & 37202 & 88681 & 45505 & 56756\\
    \arrayrulecolor{black}\bottomrule
    \end{tabular}}
    }
    \end{center}
    \end{minipage}
\end{table}

Firstly, introducing RESs in the Renew1 and Renew2 configurations results in decreased internal production costs and lower power outputs due to reduced available production compared to the Basecase configuration. In both the Renew1 and Renew2 configurations, despite lower internal generation that can be sold in the day-ahead market, the RL agent learns to bid strategically to keep the MCP high and take advantage of the increased price difference between the MCP and the PMC of the VPP, yielding a significant increase in its net profits. We observe that the increase in net profits follows the increase in installed RESs between the Baseline, Renew1, and Renew2 configurations.
Secondly, the RL agent must deal with the variability and stochasticity of the power generation introduced by the RESs in the Renew1 and Renew2 configurations. During the bidding process, the only available information is the (deterministic) power generation forecast of the RESs $p^\mathrm{res}_i,\forall i\in\mathcal{R}$. We observe that omitting the generation forecast (w/o FC case) from the state vector $s$ of the RL agent leads to a considerable decrease in profitability in both Renew1 and Renew2 configurations. This result can intuitively be explained by the lack of knowledge about the VPP power output, prompting the RL agent to place conservative bids that ultimately result in robust but suboptimal behavior. 

Finally, introducing a BESS in the Bat1 configuration increases the flexibility of the VPP compared to the Basecase scenario. The battery allows an increase in energy consumption in the first 6 hours of the day, which is stored and sold later when MCPs are higher. This observation suggests that the RL agent learns to strategically utilize battery flexibility, leading to additional arbitrage revenues and higher profits.

\section{Conclusion} \label{sec:concl}
This paper introduces a novel safe RL algorithm for the strategic bidding of VPPs in electricity markets. Numerical results show that the agent successfully learns a safe and highly competitive bidding strategy, as reflected by the low activation of the safety shield during testing and the increased profits of the agent compared to the price-taking bidding strategy. Additionally, we demonstrate that introducing the shield-activation penalty and incorporating predictive information, such as renewable energy forecast, significantly improves the optimality and safety of the agent's actions.
These findings demonstrate the potential of using safe RL techniques to optimize the bidding strategies of VPPs while respecting the network constraints and limits of internal components.

Our forthcoming research will focus on evaluating the proposed algorithm on larger-scale test cases, therefore testing the scalability of the approach. This effort will entail deriving computationally more efficient methodologies for designing the safety shield. Additionally, we intend to assess the suitability of more advanced deep RL techniques, such as the soft actor-critic approach, for enhancing competitive bidding.

\bibliographystyle{IEEEtran}
\bibliography{bibliography.bib}

\end{document}